\documentclass{elsart}
\usepackage{amsmath,amssymb,color,graphicx}
\journal{Physics Letters B}

\newcommand{\Sqrt}[1]{\sqrt{\mathstrut #1}}

\def\bfsm#1{\mathstrut\mbox{\scriptsize{\boldmath $#1$}}\mathstrut}

\newcommand{\ep}{\varepsilon}

\newcommand{\beq}{\begin{equation}}
\newcommand{\eeq}{\end{equation}}
\newcommand{\ba}{\begin{array}}
\newcommand{\ea}{\end{array}}
\newcommand{\bea}{\begin{eqnarray}}
\newcommand{\eea}{\end{eqnarray}}
\newcommand{\bal}{\begin{align}}  
\newcommand{\eal}{\end{align}}
\newcommand{\bi}{\begin{itemize}}  
\newcommand{\ei}{\end{itemize}}
\newcommand{\ben}{\begin{enumerate}}  
\newcommand{\een}{\end{enumerate}}

\newcommand\hide[1]{}

\newcommand{\tr}{\mathrm{tr}}

\renewcommand{\Re}{{\rm Re}\,}
\renewcommand{\Im}{{\rm Im}\,}

\newcommand{\ds}[1]{
  \setbox0=\hbox{\ensuremath{#1}}
  \hbox to\wd0{\hbox to0pt{\hbox to\wd0{\hss/\hss}\hss}\box0}}
\newcommand{\rme}{{\rm e}}
\newcommand{\rmi}{{\rm i}}
\newcommand{\rmd}{{\rm d}}
 
\newcommand{\MeV}{\,{\rm MeV}}

\newcommand{\diag}{{\rm diag}}

\begin{document}
\begin{frontmatter}
\title{Gauge dynamics in the PNJL model:\\
Color neutrality and Casimir scaling}

\author{Hiroaki Abuki}
\address{Institut f\"ur Theoretische Physik, 
 J.W.\ Goethe Universit\"at, D-60438 Frankfurt am Main, Germany}
\author{Kenji Fukushima}
\address{Yukawa Institute for Theoretical Physics,
         Kyoto University, Oiwake-cho, Kitashirakawa,
         Sakyo-ku, Kyoto 606-8502, Japan}
\begin{abstract}
 We discuss a gauge-invariant prescription to take the mean-field
 approximation self-consistently in the PNJL model
 (Nambu--Jona-Lasinio model with the Polyakov loop).  We first address
 the problem of non-vanishing color density in normal quark matter,
 which is an artifact arising from gauge-fixed treatment of the
 Polyakov loop mean-fields.  We then confirm that the gauge average
 incorporated in our prescription resolves this problem and ensures
 color neutrality.  We point out that the proposed method has an
 advantage in computing the expectation value of any function of the
 Polyakov loop matrix.  We discuss the Casimir scaling as an immediate
 application of the method.
\end{abstract}
\end{frontmatter}

\paragraph*{Introduction}

The interplay between the QCD phase transitions of chiral restoration
and color deconfinement at finite temperature and/or density has been
attracting much interest recently.  There are a lot of attempts to
describe confinement-deconfinement physics by means of effective
models in terms of the Polyakov loop~%
\cite{Gocksch:1984yk,Ilgenfritz:1984ff,Fukushima:2003fw,%
Meisinger:2003id,Dumitru:2003hp,Megias:2004hj,Ratti:2005jh,%
Ghosh:2006qh,Pisarski:2006hz,Roessner:2006xn,Fukushima:2006uv,%
Fu:2007xc,Ciminale:2007sr,Schaefer:2007pw,Blaschke:2007np,%
Ciminale:2007ei,Kashiwa:2007hw,Rossner:2007ik,Abuki:2008tx,%
Abuki:2008nm,Fukushima:2008wg,Hell:2008cc}.
One successful approach that can describe both the chiral and
deconfinement transitions (crossovers) is the Polyakov-loop augmented
Nambu--Jona-Lasinio (PNJL) model.  This model accommodates
self-consistent treatment for two approximate order parameters; the
Polyakov loop $L$ for deconfinement and the chiral 
condensate $\langle\bar{\psi}\psi\rangle$ for chiral restoration.  Here
the former works as an exact order parameter in the quenched limit
(i.e.\ $m_q\to\infty$), while the latter is exact in the chiral limit
(i.e.\ $m_q\to0$).  Due to a particular form of coupling between $L$
and $\langle\bar{\psi}\psi\rangle$
the PNJL model has a general tendency to make two crossovers in $L$
and $\langle\bar{\psi}\psi\rangle$ come close to each
other~\cite{Fukushima:2003fw}.  Besides, it has turned out that the
bulk thermodynamics resulting from the model shows remarkable
agreement with numerical data from the lattice QCD
simulation~\cite{Ratti:2005jh}.

It is known by now that pathological behavior arises from a simple
mean-field ansatz for the Polyakov loop matrix
\cite{Roessner:2006xn,Blaschke:2007np,Ciminale:2007ei,Hell:2008cc}.
In Ref.~\cite{Abuki:2008ht} one of the present authors found that a
saddle-point approximation on the Polyakov loop matrix leads to
unphysical non-zero color density even in the normal phase of quark
matter (see also Refs.~\cite{GomezDumm:2008sk,Abuki:2008iv}).  This is
not a principle problem inherent to the PNJL model but rather a
practical one associated with the mean-field approximation;  we have
to assume a certain gauge to make the color density definite, and at
the same time, for the sake of the color density computation it is
convenient to take a special gauge in which the Polyakov loop $L$ is
diagonal.  These two gauge choices are, however, not necessarily
compatible.  In other words, the color chemical potential matrix and
the Polyakov loop matrix are not commutable.

The problem comes from the fact that we need to treat the Polyakov
loop mean-field not as a traced 
quantity $\ell\equiv\frac{1}{N_c}\langle\tr L\rangle$ but as a matrix
$L(\varphi_1,\varphi_2)\equiv\diag(\rme^{\rmi\varphi_1},\rme^{\rmi\varphi_2},%
\rme^{-\rmi(\varphi_1+\varphi_2)})$
when we evaluate the color density.  Then we face another undesirable
situation.  That is, in terms of $\varphi_1$ and $\varphi_2$, it is
hard to realize a difference between the Polyakov loop $\ell$ and the
anti-Polyakov loop
$\bar{\ell}\equiv\frac{1}{N_c}\langle\tr L^\dagger\rangle$ 
which are both real numbers~\cite{Fukushima:2006uv,Allton:2005gk}.  It is
claimed in Ref.~\cite{Rossner:2007ik} that the fluctuation around the
mean-field induces a difference between $\ell$ and $\bar{\ell}$.  We
should note, however, that all these problems do not appear if we
treat $\ell$ and $\bar{\ell}$ as the relevant mean-fields, which works
unless we consider color degrees of freedom such as color
superconductivity~\cite{colorsuper}.  In fact, in the
color-superconducting phase, we cannot express the quasi-quark
contribution to the thermodynamic potential solely in terms of $\ell$
and $\bar{\ell}$, but it inevitably involves the matrix elements of
$L$~\cite{Roessner:2006xn}.

The present Letter aims to propose a resolution to circumvent these
shortcomings of the 
simple mean-field approximation.  We would
emphasize that our prescription not only improves the mean-field
approximation but also encompasses correct gauge dynamics from which
the neutrality with respect to gauge charge is derived (i.e.\
the Gauss law).

\paragraph*{Model and mean-field approximation}

We here explain our model, the ingredients of which are the
Polyakov-loop matrix model~\cite{Dumitru:2003hp,Pisarski:2006hz} and
the NJL model~\cite{Hatsuda:1994pi}.  The difference from the PNJL
model lies in a mean-field evaluation for the Polyakov loop
\cite{Gocksch:1984yk,Fukushima:2003fw,Megias:2004hj,Kogut:1981ez}.

First, let us address the pure gluonic sector.  We assume that the
pure gluonic dynamics would be described by the nearest neighbor
interaction of the traced Polyakov loop as
\beq
 S_{\mathrm{g}}[L]= -N_c^2\,\rme^{-a/T} \sum_{\vec{x},\hat{n}}
  l(\vec{x})\, l^\ast(\vec{x}+\hat{n}),
\label{eq:matrixmodel}
\eeq
with $l\equiv\frac{1}{N_c}\tr L$ and
$l^\ast\equiv\frac{1}{N_c}\tr L^\dagger$.  
This form of the simplest matrix model~\cite{Gupta:2006qm} is to be
postulated from the leading-order contribution in the strong coupling
expansion, which specifies the $T$-dependent interaction strength 
with $a$ being a model parameter~\cite{Fukushima:2003fw}.

The action (\ref{eq:matrixmodel}) looks like a spin model.  We
then make use of the Weiss approximation with the neighboring spin
sites treated as the mean-fields.  Hence, the mean-field action is
\beq
  S_{\mathrm{mf}}[\alpha,\beta] = -N_c\sum_x \bigl[ \alpha\,\Re l(x)
   + \rmi\beta\,\Im l(x)\bigr],
\eeq
where $\alpha$ and $\beta$ correspond to the Polyakov loop
mean-fields.  Finite $\beta$ would be induced by $\mathcal{C}$-odd
terms at finite $\mu$.  We denote the Polyakov loop expectation
values, hereafter, as $\ell\equiv\langle l\rangle_{\mathrm{mf}}$ and
$\bar{\ell}\equiv\langle l^\ast\rangle_{\mathrm{mf}}$.  
The expectation value $\langle \dots\rangle_{\mathrm{mf}}$ refers to the
average over the Polyakov loop matrix with the mean-field action.

In the Weiss mean-field approximation the free energy is defined by%
\footnote{Although the variational principle seems to break down due
  to the sign problem at $\mu\neq0$, the saddle-point of this
  mean-field free energy leads to a good
  approximation.  See Ref.~\cite{Fukushima:2006uv} for details.}
\beq
  \frac{V}{T} f_{\mathrm{g}}(\alpha,\beta) = \bigl\langle
  S_{\mathrm{g}}[L] - S_{\mathrm{mf}}[L] \bigr\rangle_{\mathrm{mf}} - \ln\int
  \mathcal{D} L\,\rme^{-S_{\mathrm{mf}}[L]}.
\label{eq:f_g}
\eeq
It is possible to find a closed analytic expression of
$f_{\mathrm{g}}(\alpha,\beta=0)$ 
at $\mu=0$, but we have to rely on numerical calculation to evaluate
$f_{\mathrm{g}}(\alpha,\beta)$ 
for $\mu\neq0$.

We can fix the parameter $a$ by requiring that the pure gluonic theory
has a first-order phase transition of color deconfinement at
$T=270\MeV$ when $\mu=0$.  This condition results in
\beq
  a = 542\MeV.
\eeq

Next, we shall consider how to add the contribution of dynamical
quarks in the mean-field approximation.  We simply add dynamical
quarks using the quasi-quark approximation with the same Polyakov loop
coupling as the PNJL model.  
In our notation $\frac{V}{T}\Omega_{\mathrm{q}}(\sigma_i,L)$ denotes the
quark thermodynamic potential with the chiral condensates,
$\sigma_u$, $\sigma_d$, and $\sigma_s$, 
giving the total mean-field free energy,
\beq
 f_{\mathrm{mf}}(\sigma_i,\alpha,\beta) = f_{\mathrm{g}}(\alpha,\beta)
  + \bigl\langle\Omega_{\mathrm{q}}(\sigma_i;L)\bigr\rangle_{\mathrm{mf}}.
\label{eq:f_mf}
\eeq
Although $\Omega_{\mathrm{q}}$ is a complex function of $L$ 
for $\mu\neq0$, its expectation value as a function of $\alpha$ 
and $\beta$ is real.  This is because the imaginary part contributing 
to the thermodynamic potential is odd under $\mathcal{C}$, i.e.\
$L\to L^\dagger$ 
transformation.  That is,
\beq
 \begin{split}
 \bigl\langle\Omega_{\mathrm{q}}(\sigma_i,L)\bigr\rangle_{\mathrm{mf}}
  &= \frac{1}{z_{\mathrm{mf}}}\int \rmd L\, \rme^{N_c\alpha\Re l} \\
   & \qquad\times\bigl[\cos(N_c\beta\Im l)\Re\Omega_{\mathrm{q}}%
   -\sin(N_c\beta\Im l)\Im\Omega_{\mathrm{q}}\bigr],
 \end{split}
\eeq
where $z_{\mathrm{mf}}$ is the normalization given as
$z_{\mathrm{mf}}=\int\rmd L\,\rme^{N_c\alpha\Re l}\cos(N_c\beta\Im l)$,
which is manifestly real.  As for the Polyakov loop, we readily find
\begin{align}
 \ell &=
  \frac{1}{z_\mathrm{mf}}\int \rmd L\,\rme^{N_c\alpha\Re l}
  \bigl[\cos(N_c\beta\Im l)\Re l-\sin(N_c\beta\Im l)\Im l\bigr],\\
 \bar{\ell} &=
  \frac{1}{z_\mathrm{mf}}\int \rmd L\,\rme^{N_c\alpha\Re l}
  \bigl[\cos(N_c\beta\Im l)\Re l+\sin(N_c\beta\Im l)\Im l\bigr].
\end{align}
It is obvious from the above that $\ell$ and $\bar{\ell}$ are
different by the presence of the imaginary (${\mathcal C}$-odd) part
induced by $\beta\neq0$ at 
finite $\mu$~\cite{Fukushima:2006uv,Dumitru:2005ng}.

We now must specify the concrete form of
$\Omega_{\mathrm{q}}(\sigma_i,L)$.  
To this end, here, we shall augment the NJL model with the Polyakov loop
coupling (i.e.\ the PNJL model).
Then $\Omega_{\mathrm{q}}(\sigma_i,L)$ take the following form;
\beq
 \begin{split}
 \Omega_{\mathrm{q}}(\sigma_i,L) &=
  g_S(\sigma_u^2+\sigma_d^2+\sigma_s^2)+4g_D\sigma_u\sigma_d\sigma_s
  -2N_c\sum_i\int^\Lambda\frac{\rmd^3\boldsymbol{p}}{(2\pi)^3}\,
  \ep_i(\boldsymbol{p}) \\
 &\quad -2T\sum_i\sum_{\lambda=\pm1} \int^\infty
  \frac{\rmd^3\boldsymbol{p}}{(2\pi)^3} \ln\det
  \bigl(1+L^\lambda\,\rme^{-(\varepsilon_i(\bfsm{p})-\lambda\mu)/T}\bigr),
 \end{split}
\label{eq:omegaq}
\eeq
where the quasi-quark dispersion relations are
$\ep_i(\boldsymbol{p})=\sqrt{p^2+M_i^2}$
with the constituent quark masses being
  $M_u = m_u-2g_S\sigma_u-2g_D\sigma_d\sigma_s$,
  $M_d = m_d-2g_S\sigma_d-2g_D\sigma_s\sigma_u$, 
  and $M_s = m_s-2g_S\sigma_s-2g_D\sigma_u\sigma_d$.
We note that $\lambda=+1$ and $-1$ in the above are the quasi-quark
and quasi-antiquark contributions, respectively.

We take the same parameter set in the NJL part as in
Ref.~\cite{Hatsuda:1994pi};
  $\Lambda=631.4\MeV$,
  $m_u =m_d =5.5\MeV$, $m_s=135.7\MeV$,
  $g_S\Lambda^2=3.67$, 
  and $g_D\Lambda^5=-9.29$.
Then $\sigma_u=\sigma_d$ always holds due to isospin symmetry in the
strong interaction.

This model has one more parameter, that is the normalization of
$f_{\mathrm{g}}(\alpha,\beta)$.  
The right-hand side of Eq.~(\ref{eq:f_g}) is proportional to the number
of space points, $N$, and thus 
$f_{\mathrm{g}}(\alpha,\beta)\propto T\cdot N/V$ which
carries the mass dimension of the energy density.  Here, $N/V$ is a
model parameter corresponding to $b$ discussed in
Ref.~\cite{Fukushima:2008wg}.  We can fix $N/V$ by the condition that
the chiral and deconfinement crossovers take place near $T=200\MeV$.
In this way, we find
\beq
 N/V=0.02\Lambda^3.
\eeq

\paragraph*{Quark and color densities}
Once we determine the mean-fields $\{\alpha,\beta,\sigma_i\}$ by
solving the gap equations, 
$\partial f_{\mathrm{mf}}/\partial\alpha
=\partial f_{\mathrm{mf}}/\partial\beta
=\partial f_{\mathrm{mf}}/\partial\sigma_i =0$,
we can calculate various physical quantities.

The quark number density, i.e.\
$n_q=-\partial f_{\mathrm{mf}}/\partial\mu$,
can be expressed as
\beq
 \begin{split}
 n_q &= \frac{1}{z_{\mathrm{mf}}}\sum_i\int
  \frac{\rmd^3\boldsymbol{p}}{(2\pi)^3} \int \rmd L\,
  \rme^{N_c(\alpha\Re l+\rmi \beta\Im l)} \\
 & \quad\times \tr\biggl[\frac{1}{L^\dagger
  \rme^{(\ep_i(\boldsymbol{p})-\mu)/T}+1}-
  \frac{1}{L\,\rme^{(\ep_i(\boldsymbol{p})+\mu)/T}+1}\biggr].
 \end{split}
\label{eq:q_density}
\eeq
Next we consider color densities.  Since the phase $A_4$ of the
Polyakov loop matrix, $L=\exp[\rmi A_4/T]$, could be regarded as the
color chemical potential, the color density is then given by
differentiating the integrand of $f_{\mathrm{mf}}$ with respect to
$(-\rmi A^a_4)$.
This leads to
\beq
 \begin{split}
 n_a &= \frac{1}{z_{\mathrm{mf}}} \sum_i\int
  \frac{\rmd^3\boldsymbol{p}}{(2\pi)^3} \int\rmd L\,
  \rme^{N_c(\alpha\Re l+\rmi\beta\Im l)} \\
 &\quad\times \tr \biggl[\frac{1}{L^\dagger
  \rme^{(\ep_i(\boldsymbol{p})-\mu)/T}+1} \,T_a - \frac{1}{L\,
  \rme^{(\ep_i(\boldsymbol{p})+\mu)/T}+1} \,T_a \biggr].
 \end{split}
\label{eq:density}
\eeq
Here $T_a$'s are the $\mathrm{SU}(N_c)$ algebra in the fundamental
representation.  In deriving this we have made use of the cyclicity in
the trace.  The group integration in Eq.~(\ref{eq:density}) is hard to
perform in general.  Usually we take the Polyakov gauge in which $L$
is diagonal with two angle variables $\varphi_1$ and $\varphi_2$, and
then we can express $\rmd L$ as $\rmd\varphi_1\rmd\varphi_2$
accompanied by the $\mathrm{SU}(N_c=3)$ Haar measure
$\mu(\varphi_1,\varphi_2)\equiv[\sin(\varphi_1\!-\!\varphi_2)
+\sin(2\varphi_1\!+\!\varphi_2)
+\sin(\varphi_1\!+\!2\varphi_2)]^2/(3\pi^2)$.
This procedure works straightforwardly for Eq.~(\ref{eq:q_density})
but not for Eq.~(\ref{eq:density}) because of the presence of $T_a$.
The color density is gauge dependent, however, so we should fix
the gauge to define this quantity.
If we take the Polyakov gauge as usual, then the color density for
only the $T_3$ and $T_8$ components (belonging to the Cartan subalgebra
of $\mathrm{SU}(3)$) have non-vanishing integrands.
We then can define the red, green, and blue quark densities as
$n_r=\frac{1}{3}n_q+\frac{1}{2}n_3+\frac{1}{2\sqrt{3}}n_8$,
$n_g=\frac{1}{3}n_q-\frac{1}{2}n_3+\frac{1}{2\sqrt{3}}n_8$, and
$n_b=\frac{1}{3}n_q-\frac{1}{\sqrt{3}}n_8$.

\paragraph*{Approximation}
Here we introduce an approximation which greatly reduces the
computational cost.  That is,
\beq
 \bigl\langle\ln\det(1+L\,\rme^{-(\ep_i-\mu)/T})\bigr\rangle_{\mathrm{mf}}
 \to\ln\bigl\langle\det(1+L\,\rme^{-(\ep_i-\mu)/T})\bigr\rangle_{\mathrm{mf}},
\eeq
and the same for the antiquark part.  With this approximation applied
to the free energy expression, we can reduce the three-dimensional
integral with respect to $\{\varphi_1,\varphi_2,p\}$ to the
one-dimensional $p$-integral with given $\ell$ and $\bar{\ell}$ which
result from the integral over $\{\varphi_1,\varphi_2\}$ independently
of $p$.  We have numerically confirmed that this approximation works
excellently well.

\paragraph*{Standard PNJL model treatment}

For comparison to the simple mean-field approximation used in
literature~\cite{Roessner:2006xn}, we shall calculate the same
physical quantities using the standard PNJL model,
\begin{align}
 \tilde{f}_{\mathrm{mf}}(\sigma_i,L) &=
  V_{\mathrm{glue}}[l,\bar{l}] \notag\\
 & +g_S(\sigma_u^2+\sigma_d^2+\sigma_s^2)
   +4g_D\sigma_u\sigma_d\sigma_s-2N_c\sum_i
   \int^\Lambda \frac{\rmd^3\boldsymbol{p}}{(2\pi)^3}\,
   \ep_i(\boldsymbol{p}) \notag\\
 & -2T\sum_i\sum_{\lambda=\pm1}\int^\infty
  \frac{\rmd^3\boldsymbol{p}}{(2\pi)^3} \ln\det
  \bigl(1\!+\!L^\lambda\,
  \rme^{-(\ep_i(\boldsymbol{p})-\lambda\mu)/T}\bigr),
\label{eq:simple}
\end{align}
where
\beq
  V_{\mathrm{glue}}[l,\bar{l}]=-bT\Bigl\{54\,\rme^{-a/T}
  l\,\bar{l} + \ln\bigl[ 1-6\,l\,\bar{l}-3(l\,\bar{l})^2
  +4(l^3+\bar{l}^3) \bigr]\Bigr\}
\eeq
with $a=664\MeV$, $b=0.026\Lambda^3$ so that the transition (crossover)
temperatures with and without dynamical quarks are $270\MeV$ and $200\MeV$ 
respectively, as explained previously.  We will refer to this model as
the \textit{standard} PNJL model hereafter.

The above expression 
for $\tilde{f}_{\mathrm{mf}}(\sigma_i,L)$ is given
in terms of the gauge invariant mean-fields, $\ell$ and $\bar{\ell}$, so
that one can 
evaluate them without difficulty to find $\bar{\ell}>\ell$ at
non-zero $\mu$.  The serious problem arises when we are interested in
quantities associated with color degrees of freedom.  We cannot
express the color density, $n_a$, using $\ell$ and $\bar{\ell}$ only,
as seen from Eq.~(\ref{eq:density}).   That is also the case if the
color-superconducting phase is considered in the PNJL model.

Practically, in such a situation, one may well assume the 
mean-fields, $\varphi_1$ and $\varphi_2$ (or $\phi_3$ and $\phi_8$), to
characterize the Polyakov loop matrix as
$L=\diag(\rme^{\rmi\varphi_1},\rme^{\rmi\varphi_2},%
\rme^{-\rmi(\varphi_1+\varphi_2)})
=\exp[\rmi(\phi_3 T_3+\phi_8 T_8)/T]$.  
This prescription is quite problematic, however, though adopted
frequently.  The traced Polyakov loops, $\ell$ and $\bar{\ell}$, become
complex with non-zero $\phi_3$ and $\phi_8$ in general.  To avoid this
artifact, one could assume $\phi_8=0$, but such an assumption is not
consistent with $\bar{\ell}\neq \ell$ at finite $\mu$.  More seriously
an unphysical color density is induced by the mean-fields, which is, of
course, an artifact of this prescription.
One may want to cancel the color density by introducing
color chemical potentials \cite{Abuki:2008ht,GomezDumm:2008sk},
but as soon as one does so, another undesirable problem seems to come
out immediately \cite{Blaschke}.

We define the magnitude of the color density by
$n_c=(\sum_{a=1}^8n_a^2)^{1/2}$~\cite{Abuki:2008ht}.  
This quantity is invariant under gauge rotation. 
Thus, under the assumption $\phi_8=0$, the color density magnitude is
$n_c=\frac{2}{\Sqrt{3}}|n(\phi_3)-n(0)|$, 
where we define
\beq
  n(\phi_3)=2\sum_i \int\frac{\rmd^3\boldsymbol{p}}{(2\pi)^3}\,
   \Re\biggl[\frac{1}{\rme^{(\ep_i(\boldsymbol{p})-\mu-\rmi\phi_3)/T}+1}
   -\frac{1}{\rme^{(\ep_i(\boldsymbol{p})+\mu+\rmi\phi_3)/T}+1}\biggr].
\eeq
We remark that $n_c$ is non-vanishing at finite $\phi_3$.  This is
simply because the phase of the Polyakov loop matrix generally has the
physical meaning of the color imaginary chemical potential, and 
so $\phi_3\neq0$ induces $n_c\neq0$.

\paragraph*{Order parameters and color densities}

\begin{figure}[t]
\centerline{
\includegraphics[width=0.48\textwidth]{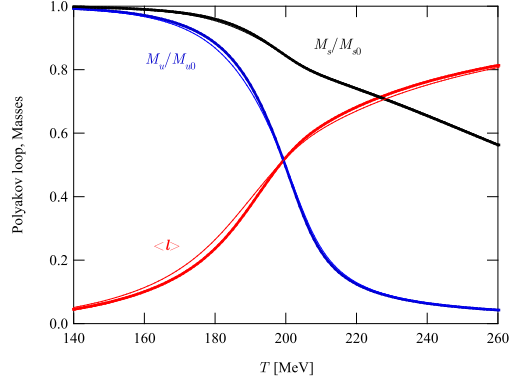}
}
\caption{The chiral and deconfinement crossovers at $\mu=0$.  The
  constituent quark masses are normalized by the vacuum 
  values, $M_{u0}$ and $M_{s0}$.  The thin lines show the results from
  the standard PNJL model given in Eq.~(\ref{eq:simple}).}
\label{fig:lock}
\end{figure}

We first show the order parameters for chiral restoration and color
deconfinement in Fig.~\ref{fig:lock} as a function of $T$ at $\mu=0$.
The thick curves represent the constituent quark masses and the
Polyakov loop obtained from Eq.~(\ref{eq:f_mf}), while the thin curves
are the results in the standard PNJL model with
Eq.~(\ref{eq:simple}).  We note that the thick and thin curves are
very close, which means that our formulation would not spoil the nice
feature established in the standard PNJL model.

\begin{figure}[t]
\centerline{
\includegraphics[width=0.48\textwidth]{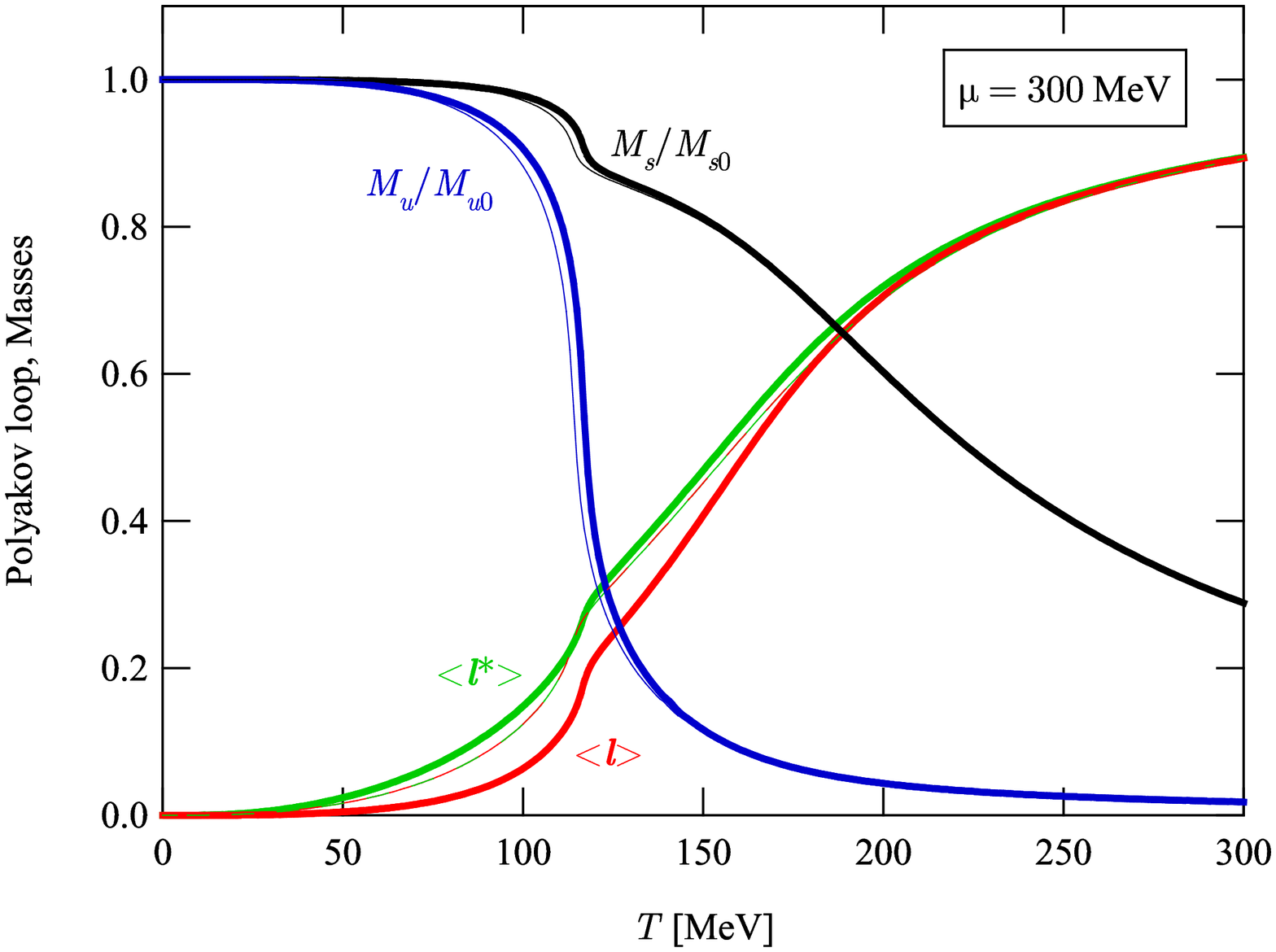}
\includegraphics[width=0.48\textwidth]{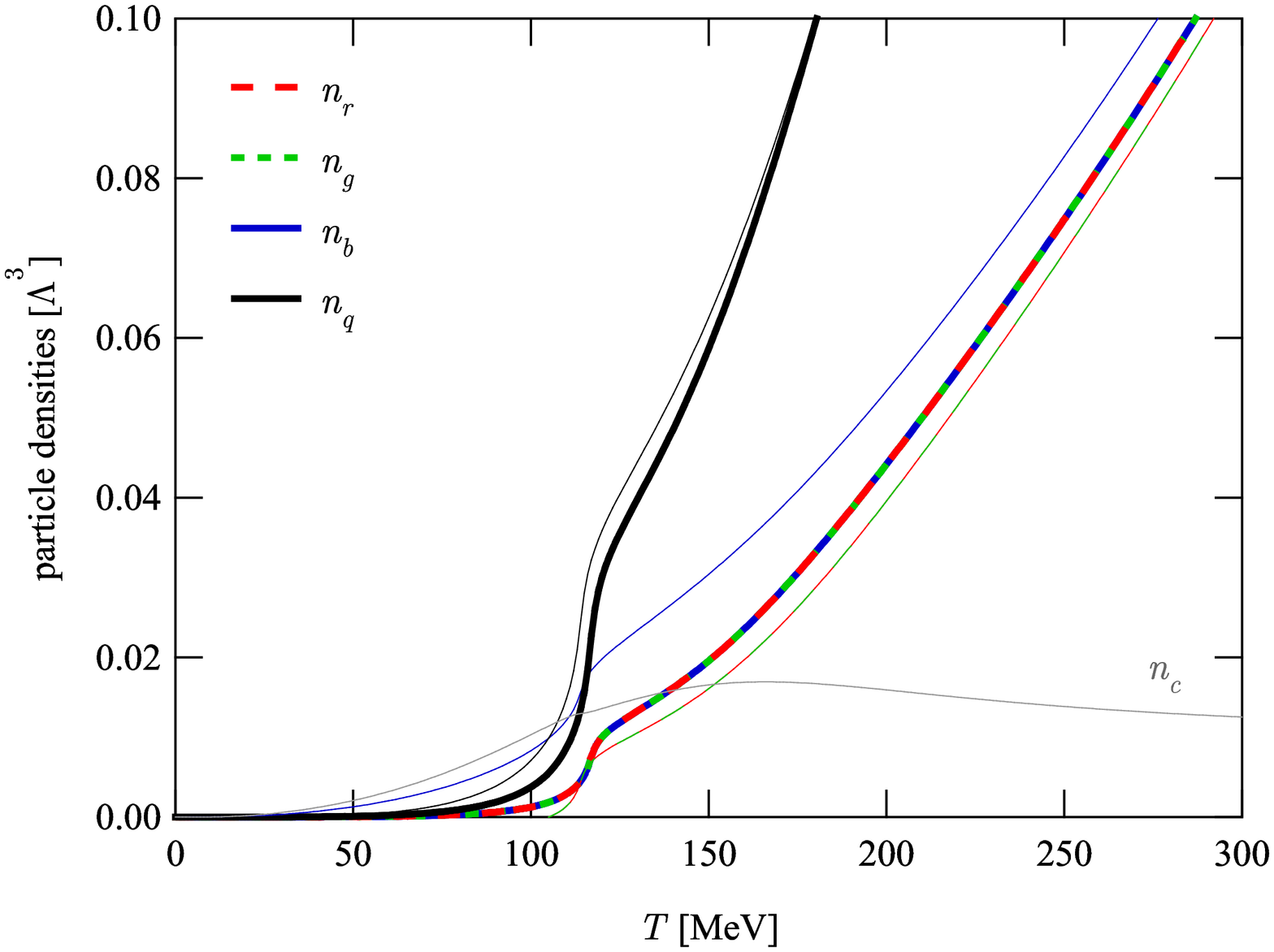}
}
\caption{Left: Constituent quark masses, $M_u$ and $M_s$, Polyakov
  loops, $\ell$ and $\bar{\ell}$, as a function of $T$ at $\mu=300\MeV$.  
  The thin curves are the results from the standard PNJL model.  Right:
  Color densities $\{n_r,n_g,n_b\}$ and the quark number density $n_q$
  as a function of $T$.  The net color density
  $n_c=|n_r+n_g-2n_b|/\sqrt{3}$ 
  is indicated by the thin line with the label $n_c$ in the case of the
  standard PNJL model. In our prescription $n_c$ is always zero.}
\label{fig:mu300}
\end{figure}

Now we shall move on to the finite density case.  In
Fig.~\ref{fig:mu300} we display the physical quantities as a function
of $T$ at $\mu=300\MeV$.  The left figure shows the constituent quark
masses and the Polyakov loop in the same way as in
Fig.~\ref{fig:lock}.  The thin lines are the results from the standard
PNJL model again.  The right figure shows the quark number density $n_q$ 
and the color densities $\{n_r,n_g,n_b\}$, where $n_q=n_r+n_g+n_b$
should be fulfilled.  We see that the thick curves have significant
difference from the thin curves resulting from the standard PNJL model.
In the case of our prescription we have $n_r=n_g=n_b$, meaning that
$n_c=|n_r+n_g-2n_b|/\sqrt{3}=0$, 
while $n_c$ is non-vanishing in the standard PNJL model as depicted by
the thin line with the label $n_c$.  It is also notable that, despite
drastic difference in the color densities, the quark number density, $n_q$, 
hardly changes; the thin line stays close to the thick line for $n_q$.

\paragraph*{Casimir scaling at finite density}

As already noted, the mean-field approximation discussed here enables
us to compute not only the traced Polyakov loop in the fundamental
representation but also the expectation value of arbitrary functions
of the Polyakov loop.  We shall take a close look at the Polyakov loop
in the higher representations as an immediate application.

The Polyakov loop in the higher representations is of special interest
with regard to the Casimir scaling hypothesis~\cite{Bali:2000un},
which may provide a crucial key to 
understanding non-perturbative aspects
of QCD such as confinement~\cite{Gupta:2006qm}.  The Casimir scaling
hypothesis claims that the color singlet potential between static
color sources in the representation $r$ is proportional to the Casimir
invariant $C_2(r)$.  The statement is rather obvious in the
perturbative regime,  but it is quite non-trivial at large distances.
From the theoretical perspective this hypothesis is 
verified up to two-loop order in the lattice perturbation theory both
in pure gauge theory~\cite{Schroder:1998vy} and in QCD with massless
dynamical quarks~\cite{Bali:2002wf}.  Beyond two-loop order the
Casimir scaling can be violated, though the violation is
tiny~\cite{Schroder:1998vy}.

\begin{table}
\begin{center}
\begin{tabular}{lllllll}
\hline\hline
$r$ & $(p,q)$ & $z^t$ & $C_2(r)$ & $d_r$ & \multicolumn{2}{c}{direct
 product expression of $V_r\equiv D(r)l_r$}
 \\ \hline
$3$ & $(1,0)$ & $z$ & $4/3$ & $1$ & $V_3=\tr L$ & \\
$\bar{3}$ & $(0,1)$ & $z^*$ & $4/3$ & $1$ & $V_{\bar{3}}=\tr{L^\dagger}$ & \\ 
$6$ & $(2,0)$ & $z^*$ & $10/3$ & $2.5$ & $V_6=(V_3^2-V_{\bar{3}})$ &\\
$8$ & $(1,1)$ & $1$ & $3$ & $2.25$ & $V_8=(|V_3|^2-1)$ &
 $(\Im V_8=0)$\\
$10$ & $(3,0)$ & $1$ & $6$ & $4.5$ & $V_{10}=(V_3V_6-V_8)$ & \\
$15_{\mathrm{a}}$ & $(2,1)$ & $z$ & $16/3$ & $4$ &
 $V_{15_a}=(V_{\bar{3}}V_6-V_3)$ & \\
$15_{\mathrm{s}}$ & $(4,0)$ & $z$ & $28/3$ & $7$ &
 $V_{15_s}=(V_3V_{10}-V_{15_a})$ & \\
$24$ & $(3,1)$ & $z^*$ & $25/3$ & $6.25$ &
 $V_{24}=(V_{\bar{3}}V_{10}-V_6)$ &\\
$27$ & $(2,2)$ & $1$ & $8$ & $6$ & $V_{27}=(|V_6|^2-V_8-1)$ &$(\Im V_{27}=0)$\\
\hline\hline
\end{tabular}
\end{center}
\vspace*{3mm}
\caption{Group theoretical factors in various representations; $r$ is
  the representation labeled by its dimension $D(r)$, $(p,q)$ is the
  corresponding weight factor, $t=p-q$ modulo $3$ is the triality,
  $z\equiv\rme^{\rmi2\pi/3}$ 
  is an element of $Z_3$, $C_2(r)$ is the quadratic Casimir invariant,
  and $d_r$ defines the ratio $d_r\equiv C_2(r)/C_2(3)$.  The dimension
  is given by $D(r)=(p+1)(q+1)(p+q+2)/2$.  In the triality zero
  representation (i.e.\ $z^{t}=1$), the Polyakov loop is insensitive to
  center symmetry, and thus it does not serve as an order parameter of
  deconfinement.}
\label{table:reps}
\end{table}

The scaling hypothesis is also tested numerically in the lattice
simulation.  In the $\mathrm{SU}(3)$ pure gauge theory at $T=0$ the
hypothesis has been verified up to the string breaking
distance~\cite{Bali:2000un}.   The Casimir scaling hypothesis also
brings a strong constraint on the Polyakov loop expectation value; the
traced Polyakov loop in any representation $r$ should satisfy the
following scaling relation irrespective of the renormalization of the
Polyakov loop~\cite{Gupta:2006qm};
\beq
 \ell_{r}^{1/d_{r}} \approx \ell_3.
\label{eq:scaling}
\eeq 
Here $\ell_{3}=\ell$ is the Polyakov loop in the fundamental
representation, $d_r$ is the ratio of the quadratic Casimir invariant;
$d_r\equiv C_2(r)/C_2(3)=\frac{3}{4}C_2(r)$.  
The relation (\ref{eq:scaling}) between the Polyakov loops in different
representations actually provides a useful test for the hypothesis,
and in Ref.~\cite{Gupta:2006qm} this test has been extensively
performed with use of lattice QCD data both in pure gauge theory and
in $N_f=2$ QCD.\ \ It has been found that the scaling violation is
visible only in the very vicinity of the first-order phase transition
in pure gauge theory, while the deviation from the scaling is more
evident in $N_f=2$ QCD particularly below the crossover temperature.
The scaling (\ref{eq:scaling}) is almost exact at high temperature
where $\ell$ is substantially large.

\begin{figure}[t]
\centerline{
\includegraphics[width=0.48\textwidth]{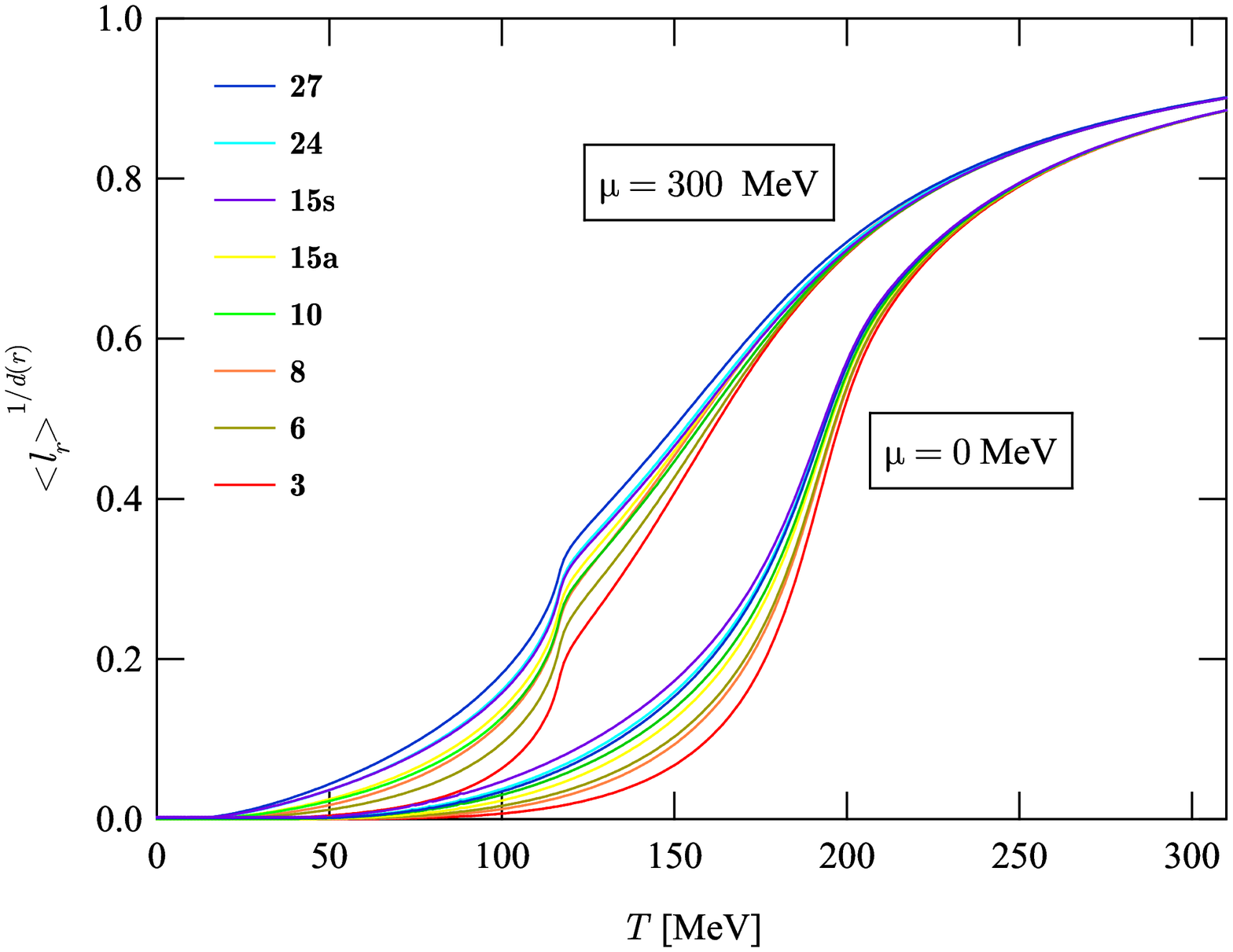}
\includegraphics[width=0.50\textwidth]{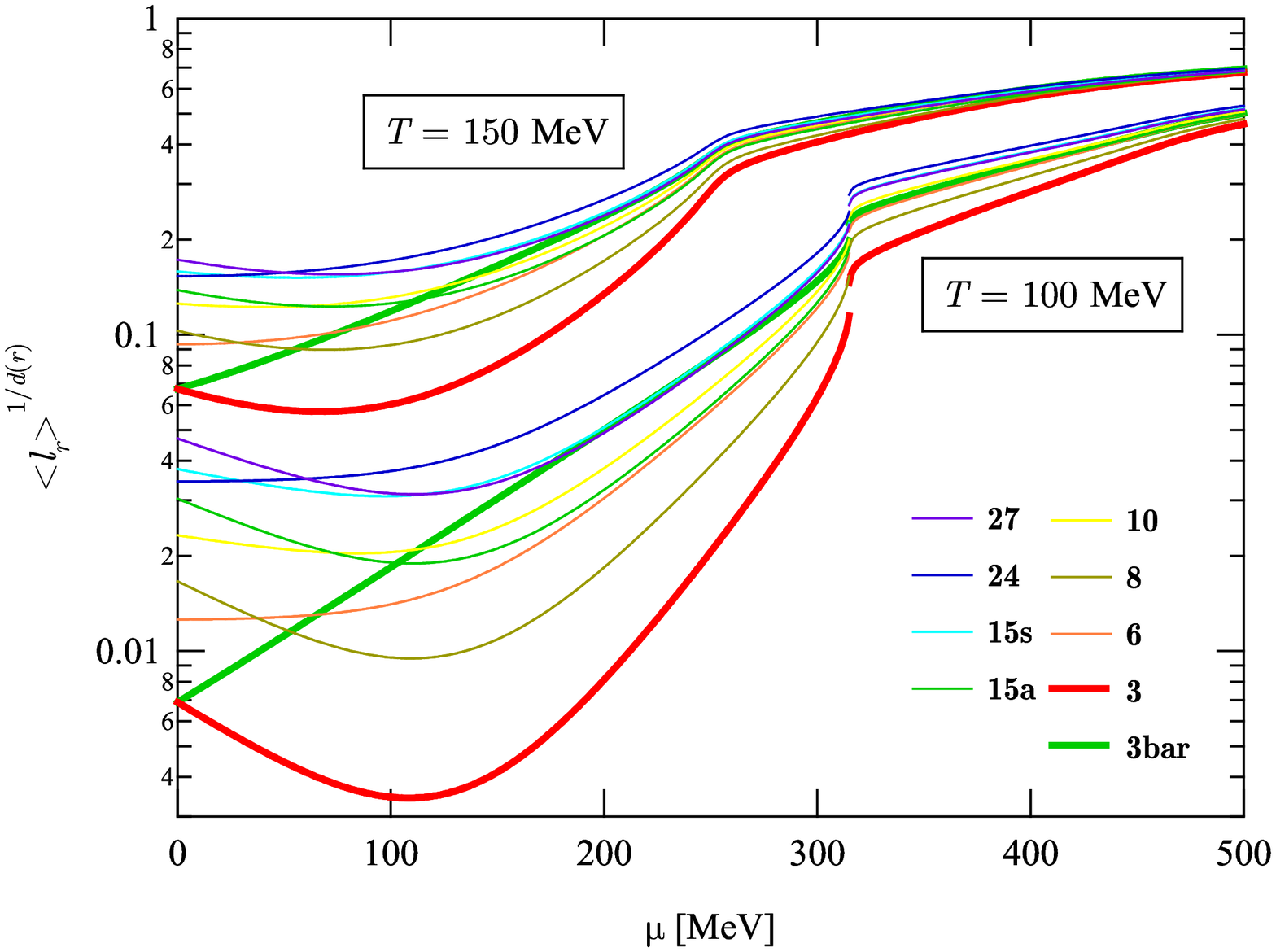}
}
\caption{Left: Scaled Polyakov loop $\ell_r^{1/d_r}$ as a function 
  of $T$ for $\mu=0$ and for $\mu=300\MeV$ in various representations
  with $d_r=C_2(r)/C_2(3)$ being the ratio of the quadratic Casimir
  invariant.  Right: Scaled Polyakov loop as a function of $\mu$ 
  for $T=100\MeV$ and for $T=150\MeV$.  It should be noted that the
  vertical axis is logarithmic.}
\label{fig:casimirscaling}
\end{figure}

It has been reported recently that the Casimir scaling of the Polyakov
loop in the fundamental (\textbf{3}) and adjoint (\textbf{8})
representations is well realized in the PNJL model at
$\mu=0$~\cite{Tsai:2008je}.  
We here present the first systematic model study on the Casimir scaling
at non-zero chemical potential.  We compute the Polyakov loop in various
representation from \textbf{3} to \textbf{27} as listed in
Tab.~\ref{table:reps}.  We can construct the Polyakov loop matrix in
higher representations by the direct products of $L$ and $L^\dagger$ in
the fundamental representation using the 
Clebsch-Gordan coefficients.  In
the left of Fig.~\ref{fig:casimirscaling} we show the various Polyakov
loops as a function of $T$ at $\mu=300\MeV$ as well as at $\mu=0$.  We
see that the scaling is good at high $T$ in both cases.  The scaling
regime is reached faster in the $\mu=0$ case; the violation of the
Casimir scaling is small at the crossover around $T=200\MeV$, while 
in the case of $\mu=300\MeV$ the deviations in $\ell_r^{1/d_r}$ are
significant at the crossover temperature $T=120\MeV$. The presence of
finite density tends to enhance the scaling violation.  
At the same time we should be careful about the interpretation; in
Ref.~\cite{Gupta:2006qm} it has been shown that this kind of
matrix-based model fails in reproducing the exact Casimir scaling
$\ell_r^{1/d_r}=\ell_3$ 
observed on the lattice in the pure gauge theory.  
One comment which we should mention here is that the scaling violation
at higher representations may be attributed to the fact that we limit
ourselves to the simplest version of the matrix model in
Eq.~(\ref{eq:matrixmodel}).  It could be possible that the inclusion
of the Polyakov loops in higher representations may diminish
artificial violation of the Casimir scaling.   
For example we could consider, $l_3(\vec{x})l_6(\vec{x}+\hat{n})$,
$l_6(\vec{x}) l_6^*(\vec{x}+\hat{n})$, 
etc., in the model action, which are allowed by $Z_3$ center symmetry in
the pure gluonic sector. Moreover, as for the Polyakov loops in the
triality zero representations such as \textbf{8}, \textbf{10}, and
\textbf{27}, one may well add their arbitrary functions in the model
action.  It would be an interesting future problem to take account of
the Polyakov loop in higher representations into the matrix model
in such a way that the model preserves charge conjugation
symmetry~\cite{Dumitru:2005ng}.

In the right of Fig.~\ref{fig:casimirscaling} we show the Polyakov
loop $\ell_r$ as a function of $\mu$ for $T=100\MeV$ and for $T=150\MeV$.  
At $T=100\MeV$ the first-order chiral transition occurs 
at $\mu=315\MeV$, while chiral restoration is smooth crossover with
increasing $\mu$ when $T=150\MeV$.  Again we notice the significant
scaling violation, though the violation is exaggerated on the
logarithmic plot.  We observe in Fig.~\ref{fig:casimirscaling} that 
some $\ell_r^{1/d_r}$'s cross each other as $\mu$ increases.  We see
that the change in \textbf{6} is milder than those in \textbf{3} and
\textbf{15}.   This is reasonable, for the excitation with the 
triality $z^*$ (like \textbf{6}) should be easier than that with $z$
(like \textbf{3} and \textbf{15}) in a medium carrying the triality $z$
at $\mu>0$.

\paragraph*{Conclusion}

We showed that the pathological problems associated with a 
simple mean-field approximation in the PNJL model can be resolved by the
use of the Weiss mean-field approximation.  We explicitly demonstrated
that the color density is vanishing in the normal phase of quark
matter.  This vanishing color density is guaranteed by the integration
with respect to the Polyakov loop, which can translate into
the Gauss law resulting from the $A_4$-integration.  
We also confirmed that $\bar{\ell}>\ell$ at finite $\mu$ is naturally
realized.

Our mean-field prescription allows us to compute the expectation value
of any function of the Polyakov loop matrix, $L$ and $L^\dagger$,
easily.  As a demonstration, we computed the Polyakov loops in various
representations from \textbf{3} to \textbf{27}.  We observed that the
Casimir scaling is violated at finite $\mu$ more than at zero density,
which is quite natural.  More interestingly, the Polyakov loop with
the triality $z$ turned out to have decreasing behavior as a function
of $\mu$ as long as $\mu$ is small.  This means that the quark
excitation bearing the same triality charge with the background medium
is less favored.  It would be an interesting future work to
incorporate couplings between the Polyakov loops in higher
representations into the matrix model and investigate the scaling
violation in wider model space.

An interesting extension of the present work would be the QCD phase
structure with the diquark condensation taken into account.  It would
be of particular interest how the diquark condensates and the colored
Polyakov loop matrix are entangled in a color-superconducting medium.
The work along this line certainly deserves future investigations.

\vspace*{1em}
H.~A.\ thanks D.~Blaschke, T.~Brauner, M.~Buballa, D.~Rischke,
M.~Ruggieri and F.~Sandin for discussions.  The present work was
supported in part by the Alexander von Humboldt Foundation.  Numerical
calculations were performed using the facilities of the Frankfurt
Center for Scientific Computing.  K.~F. is grateful to D.~Rischke for
the hospitality at 
the Institut f\"ur Theoretische Physik of Johann Wolfgang
Goethe-Universit\"at Frankfurt am Main, where this work was initiated.
K.~F.\ is supported by Japanese MEXT grant No.\ 20740134 and also
supported in part by Yukawa International Program for Quark Hadron
Sciences.


\end{document}